\documentclass[conference]{IEEEtran}
\IEEEoverridecommandlockouts
\usepackage{graphicx}
\usepackage{tikz}
\usetikzlibrary{arrows.meta, positioning}
\usepackage{xcolor}
\usepackage{caption,subcaption} 
\usepackage{booktabs}

\usepackage{amsmath, amssymb, amsfonts}
\usepackage{siunitx}
\usepackage{algorithm}
\usepackage{algorithmic} 
\usepackage{cite}
\usepackage{url}

\usepackage{acronym}

\acrodef{RIR}{Room Impulse Response}
\acrodef{CD}{Cosine Distance}
\acrodef{NMSE}{Normalized Mean Square Error}
\acrodef{GAN}{Generative Adversarial Network}
\acrodef{CNN}{Convolutional Neural Network}
\acrodef{DDPM}{Denoising Diffusion Probabilistic Models}
\acrodef{PINN}{Physics-informed Neural Network}
\acrodef{EDC}{Energy Decay Curve}
\acrodef{ULA}{uniform linear array}
\acrodef{SCI}{Spline Cubic Interpolation}
\acrodef{VM}{Virtual Microphone}
\acrodef{SNR}{Signal-to-Noise Ratio}
\acrodef{SIR}{Signal-to-Interference Ratio}
\acrodef{STOI}{Speech Intelligibility}
\acrodef{SI-SDR}{Scale-Invariant Signal-to-Distortion Ratio}
\acrodef{RTF}{Relative Transfer Function}
\acrodef{ATF}{Acoustic Transfer Functions}
\acrodef{MVDR}{Minimum Variance Distortionless Response}
\acrodef{LMS}{Least Mean Squares}
\acrodef{RLS}{Recursive Least Squares}
\acrodef{STFT}{short-time Fourier transform}
\acrodef{DOA}{direction-of-arrival}
\acrodef{GSC}{Generalized Sidelobe Canceller}

\def\BibTeX{{\rm B\kern-.05em{\sc i\kern-.025em b}\kern-.08em
    T\kern-.1667em\lower.7ex\hbox{E}\kern-.125emX}}
\begin{document}

\title{On the Usefulness of Diffusion-Based Room Impulse Response Interpolation to Microphone Array Processing}

\author{\IEEEauthorblockN{Sagi Della Torre}
\IEEEauthorblockA{
\textit{Bar-Ilan University, Israel}\\
0009-0002-9685-6665
}
\and
\IEEEauthorblockN{Mirco Pezzoli}
\IEEEauthorblockA{
\textit{Politecnico di Milano, Italy}\\
0000-0003-1296-0992
}
\and
\IEEEauthorblockN{Fabio Antonacci}
\IEEEauthorblockA{
\textit{Politecnico di Milano, Italy}\\
0000-0003-4545-0315
}
\and
\IEEEauthorblockN{Sharon Gannot}
\IEEEauthorblockA{
\textit{Bar-Ilan University, Israel}\\
0000-0002-2885-170X}
}

\maketitle

\begin{abstract}
\ac{RIR} estimation is a fundamental problem in spatial audio processing and speech enhancement. 
In this paper, we build upon our previously introduced diffusion-based inpainting framework for \ac{RIR} interpolation and demonstrate its applicability to enhancing the performance of practical multi-microphone array processing tasks. Furthermore, we validate the robustness of this method in interpolating real-world \acp{RIR}. 
%
\end{abstract}

\begin{IEEEkeywords}
\textit{Diffusion models, \ac{RIR} interpolation, microphone array processing, \ac{VM}}
\end{IEEEkeywords}

\section{Introduction}
\acfp{RIR} play a fundamental role in acoustic signal processing \cite{polack1993playing}.
They characterize sound propagation in enclosed environments and underpin key tasks such as source localization, source separation, dereverberation, and speech enhancement.
However, acquiring dense and accurate \ac{RIR} measurements in real rooms is time-consuming and resource-intensive.
As an alternative, simulated \acp{RIR} \cite{allen1979image, habets2006room} are often used, yet they typically fail to capture the full complexity of real acoustic environments.

To address these limitations, numerous methods have been proposed to reconstruct or interpolate missing \acp{RIR} from sparse measurements.
Existing approaches range from geometry-based modeling techniques \cite{thiergart2013geometry, greco2026parametric, fahim2017sound, zea2019compressed} to data-driven, learning-based methods \cite{masuyama2025physics, miotello2024reconstruction}, all aiming to enhance spatial resolution without requiring exhaustive measurement campaigns.

In our previous work \cite{della2025diffusionrir}, we proposed \textit{DiffusionRIR}, a diffusion-based generative framework that treats the collection of \acp{RIR} as a grayscale image and formulates the reconstruction problem as an inpainting task. 
DiffusionRIR demonstrated strong performance on simulated datasets, outperforming standard interpolation baselines.
However, the evaluation in \cite{della2025diffusionrir} was limited to synthetic data, leaving open the question of how well the model generalizes to real, measured \acp{RIR}.

Beyond \ac{RIR} reconstruction itself, the quality and density of spatial acoustic information have a direct impact on downstream array signal processing applications. 
Beamforming performance, including that of the \ac{MVDR} beamformer, depends directly on the number and spatial distribution of available microphones. 
To improve performance in scenarios with limited sensing, the \acf{VM} framework  \cite{katahira2016nonlinear, yamaoka2019cnn, segawa2024neural} synthetically increases the number of channels—often using geometric assumptions and interpolation in the \ac{STFT} domain—thereby enabling more effective spatial filtering with a small number of physical microphones.

When microphone signals are available, \ac{MVDR} beamforming can be implemented using either a \ac{DOA}-based or an \ac{RTF}-based steering vector.
Previous studies \cite{shmaryahu2022importance, gannot2017consolidated} have shown that \ac{RTF}-based beamformers are effective for interference suppression; however, they inherently preserve the reverberant component captured at the reference microphone, thereby limiting dereverberation performance.
In contrast, beamforming based on the full \acp{ATF} enables not only spatial filtering but also suppression of reverberation effects, since the complete acoustic transfer function is incorporated into the steering vector. While estimating the \acp{DOA} or \acp{RTF} constitutes a non-blind problem, estimation of the full \acp{ATF} from measured microphone signals is inherently blind and significantly more challenging.
Furthermore, acquiring dense and accurate \acp{ATF} (or equivalently, \acp{RIR}) measurements, required for high-resolution beampattern synthesis, is difficult in practical acoustic environments.
This challenge motivates the reconstruction of missing \acp{RIR} from a limited set of measurements.

In this work, we assume that microphone signals are available at all array positions, while \ac{RIR} measurements are available only at a subset of microphones. 
DiffusionRIR is used to reconstruct the missing \acp{RIR}, which are then employed for beamforming. 
We assess the usefulness of DiffusionRIR for beamforming performance and evaluate its robustness using real-world acoustic measurements, thereby extending prior validation beyond synthetic data.

\section{Problem Formulation}

The objective of this work is to exploit reconstructed \acp{RIR} for spatial speech enhancement and separation. 
We build upon our previously proposed framework \cite{della2025diffusionrir}, which addressed the task of estimating missing \acp{RIR} in microphone arrays. 

Consider an array of $N$ microphones, of which only $M$ \acp{RIR} are measured. 
The remaining $L = N - M$ \acp{RIR} are unmeasured. 
Let $\mathbf{H} \in \mathbb{R}^{K \times N}$ denote the complete \ac{RIR} matrix, where each column corresponds to the impulse response at a microphone location, truncated to $K$ samples. 
The available measurements form a submatrix denoted $\mathbf{H}_{\text{measured}} \in \mathbb{R}^{K \times M}$, while the missing measurements submatrix denoted by $\mathbf{H}_{\text{missing}} \in \mathbb{R}^{K \times L}$. 
The reconstruction task can be expressed as $\mathbf{\hat{H} = \mathcal{F}(H_{\text{measured}}, \mathcal{M}),}$
where $\mathcal{M}$ is a column-wise binary mask indicating available and missing microphones, and $\mathcal{F}(\cdot)$ is the reconstruction operator. 

The resulting matrix $\hat{\mathbf{H}}$ provides a dense set of \acp{RIR} across the array. 
Beyond the reconstruction task itself, we examine the usefulness of these reconstructed \acp{RIR} in classical microphone array processing problems. 
In particular, we consider the \ac{MVDR} beamformer under the assumption that signals from all $N$ microphones are available, while \acp{RIR} are measured at only $M$ locations. 
The reconstructed \acp{RIR} are subsequently used to complement the measured responses, with the objective of enhancing beamforming performance.
%
%
%
%
\section{\ac{RIR} Interpolation using Inpainting}
In this section, we briefly summarize the DiffusionRIR framework introduced in our previous work \cite{della2025diffusionrir}, which serves as the basis for the interpolation method used throughout this study.
Reconstruction is treated as an inpainting task, drawing inspiration from recent advances in diffusion models, in particular the RePaint strategy \cite{lugmayr2022repaint}. 
The \ac{RIR} set is first recast into a grayscale image-like structure, where each column corresponds to one microphone response truncated to $K$ samples. 
Missing microphones are encoded by a column-wise binary mask $\mathcal{M}$, which ensures that measured data remain unchanged while only the unmeasured positions are estimated.

The generative diffusion process starts from Gaussian noise and iteratively refines the \ac{RIR} image via a learned reverse-diffusion operator $\mathcal{D}_\theta$. We employ OpenAI's \ac{DDPM} architecture\footnote{\url{https://github.com/openai/guided-diffusion}} with modifications for \ac{RIR}-matrix images. 
To improve precision in the low-energy parts of the \acp{RIR}, the image is split into overlapping $64 \times 64$ patches, normalized to the range $[-1,1]$, and reassembled into $\mathbf{\hat{H}}$, thereby preserving spatial and temporal consistency across the array.  

The complete procedure is summarized in Algorithm~\ref{alg:diffusionrir}.

\begin{algorithm}
\caption{DiffusionRIR: RIR Reconstruction via Diffusion Models}
\label{alg:diffusionrir}

\textbf{Input:} Measured \acp{RIR} $\mathbf{H}_{\text{measured}} \in \mathbb{R}^{K \times M}$, 
microphone mask $\mathcal{M}$ (1 = measured, 0 = missing), 
diffusion model $\mathcal{D}_\theta$ \\
\textbf{Output:} Reconstructed \ac{RIR} matrix $\mathbf{\hat{H}} \in \mathbb{R}^{K \times N}$

\begin{algorithmic}[1] 
\STATE Convert $\mathbf{H}_{\text{measured}}$ into grayscale image-like matrix $\mathbf{X}$
\STATE Split $\mathbf{X}$ into overlapping patches $\mathbf{x}$ of size $64 \times 64$
\STATE Normalize each patch into range $[-1,1]$
\FOR{each patch $\mathbf{x}$}
    \STATE Initialize $\mathbf{x}_T \sim \mathcal{N}(0, \mathbf{I})$ \hfill \COMMENT{random Gaussian noise}
    \FOR{$t = T, T-1, \dots, 1$}
        \STATE Predict noise $\epsilon_\theta(\mathbf{x}_t, t, \mathcal{M})$ using $\mathcal{D}_\theta$
        \STATE Estimate clean patch $\mathbf{x}_{t-1}$ via reverse diffusion
        \STATE Enforce consistency with measured data in $\mathcal{M}$
    \ENDFOR
    \STATE Store reconstructed patch $\mathbf{\hat{x}} = \mathbf{x}_0$
\ENDFOR
\STATE Undo patch normalization to restore amplitude scale
\STATE Reassemble patches into complete image $\mathbf{\hat{X}}$
\STATE Convert $\mathbf{\hat{X}}$ back into matrix $\mathbf{\hat{H}} \in \mathbb{R}^{K \times N}$
\RETURN $\mathbf{\hat{H}}$
\end{algorithmic}
\end{algorithm}

\section{Array Processing based on Interpolated \acp{RIR}}
Given the \acp{RIR} from the source to all microphones, one can design beamformers that exploit both spatial and temporal information to enhance the desired source and suppress reverberation and noise. 
Unlike approaches based solely on \ac{RTF}, the availability of full \ac{RIR} enables direct recovery of the transmitted signal, largely free from reverberant distortions.

In our framework, we assume that all microphone signals are available, whereas the corresponding \ac{RIR} measurements are available only at a subset of microphones. 
The missing \acp{RIR} are first reconstructed using DiffusionRIR, resulting in an estimated \ac{RIR} matrix $\hat{\mathbf{H}} \in \mathbb{R}^{K \times N}$ that spans both measured and interpolated microphone locations. 
The reconstructed spatial information is then incorporated into conventional array processing algorithms for speech enhancement and source separation.

\begin{figure}[ht]
\vspace{-6pt}
  \centering
  \includegraphics[width=\linewidth]{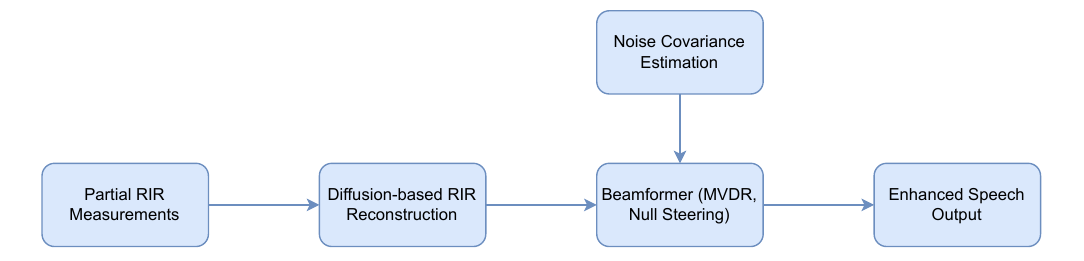}
 \addtolength{\belowcaptionskip}{-8pt}
\addtolength{\abovecaptionskip}{-8pt}
  \caption{System block diagram: Diffusion-based \acp{RIR} reconstruction followed by beamforming for speech enhancement.}
  \label{fig:system_diagram}
\end{figure}

Figure~\ref{fig:system_diagram} illustrates a block diagram of the proposed system. 
The pipeline consists of: 
(i) acquisition of partial \ac{RIR} measurements, 
(ii) diffusion-based reconstruction of the missing \acp{RIR}, 
(iii) construction of steering vectors and beamformer weights using the reconstructed \acp{RIR} together with noise covariance estimated from the available microphone signals, and 
(iv) speech enhancement through \ac{MVDR} applied to the available microphone signals (signals are not shown in the figure).

\subsection{\ac{ATF} Based MVDR Beamformer} 
Here, we focus on the \ac{MVDR} beamformer \cite{gannot2017consolidated}, whose weight vector is given by
$\mathbf{w}_{\text{{MVDR}}} = \frac{\boldsymbol{\Phi}_{nn}^{-1} \mathbf{d}}{\mathbf{d}^H \boldsymbol{\Phi}_{nn}^{-1} \mathbf{d}}$,
where $\mathbf{d}$ denotes the steering vector associated with the target source, derived from the reconstructed \acp{RIR}, and $\boldsymbol{\Phi}_{nn}$ is the noise covariance matrix estimated from noise-only segments using all available microphone signals.

By exploiting $\hat{\mathbf{H}}$, the steering vectors incorporate all \acp{RIR}, both measured and reconstructed.
Without interpolation, an \ac{ATF}-based \ac{MVDR} beamformer can only be constructed from the $M$ measured microphones, yielding a steering vector $\mathbf{d} \in \mathbb{C}^{M}$ and an $M\times M$ noise covariance matrix estimated from the same subset of channels.

After interpolation, the steering vector expands to $\mathbf{d} \in \mathbb{C}^{N}$, enabling the beamformer to leverage all available microphone signals in its design.
In the following, we examine how this extension affects beamforming performance with respect to interference suppression and dereverberation. We compare \ac{MVDR} beamforming performance under varying \ac{RIR} availability to assess the contribution of reconstructed responses.

\vspace{1pt}
\noindent \textbf{Experiment Setup:} Experiments were conducted using simulated data generated with the Pyroomacoustics package.\footnote{\url{https://pyroomacoustics.readthedocs.io/en/pypi-release/}}
A \ac{ULA} with $N=16$ microphones and an inter-element spacing of $\SI{4}{\centi\meter}$ was used, yielding an aperture of $\SI{0.64}{\meter}$. 
The room dimensions were $6 \times 5.5 \times \SI{2.8}{\meter}$, with a reverberation time of $T_{60}=\SI{300}{\milli\second}$. 
Signals were sampled at $\SI{8}{\kilo\hertz}$, and each \ac{RIR} was truncated to $2048$ samples ($\SI{256}{\milli\second}$). 
The speech source was located $2$~m in front of the array at broadside, and $7$-seconds utterances were employed.

To evaluate robustness in noisy conditions, two spatial noise scenarios were considered: 
(i) a directional noise source placed at a different room location, $2$~m from the array, and 
(ii) diffuse noise generated using a diffuse noise simulator.\footnote{\url{https://www.audiolabs-erlangen.de/fau/professor/habets/software/noise-generators}}
All noise signals were generated as pink noise, with SNR levels ranging from $-10$ to $\SI{10}{\decibel}$. 
Additionally, spatially white noise was added to the mixture at an SNR of $\SI{10}{\decibel}$.

\begin{figure}[t!]
 \centerline{\framebox{
 \includegraphics[width=7.8cm]{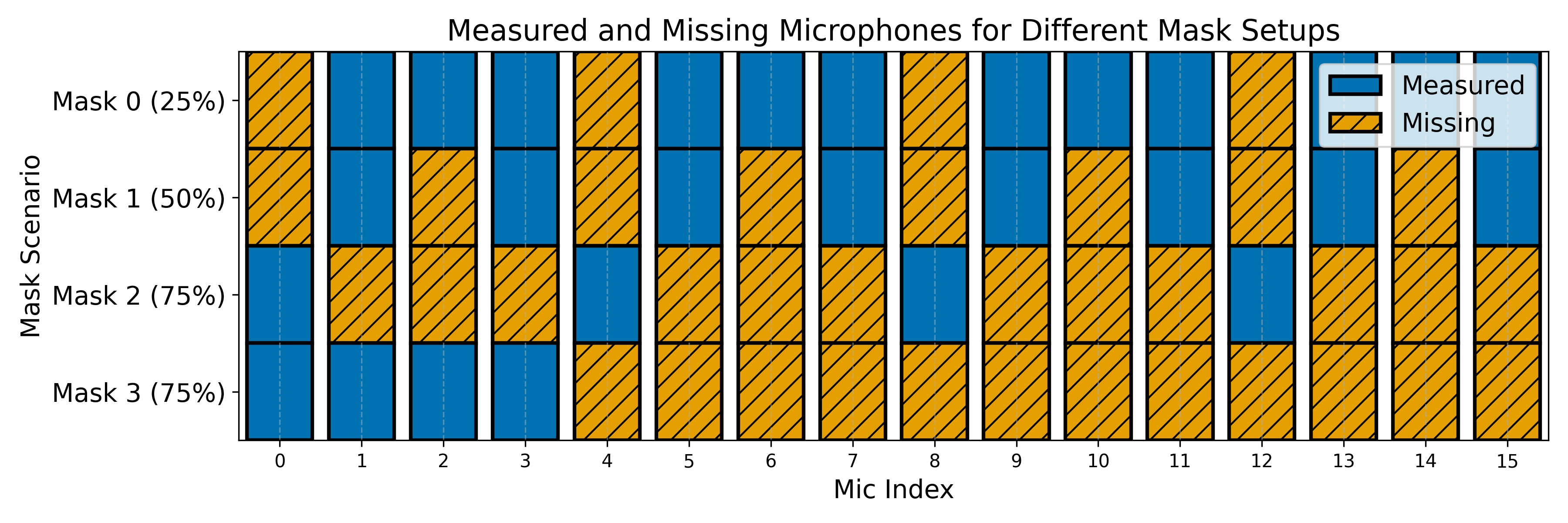}}}
 \caption{Measured (blue) and missing (orange) microphones.}
 \label{fig:masks_setup}
\end{figure}

Partial \ac{RIR} availability was determined using four masking configurations:
\begin{itemize}
    \item \textbf{Mask 0:} $12$ microphones measured, $4$ missing.
    \item \textbf{Mask 1:} $8$ microphones measured, $8$ missing.
    \item \textbf{Mask 2, 3:} $4$ microphones measured, $12$ missing.
\end{itemize}
All mask configurations are illustrated in Fig.~\ref{fig:masks_setup}.

For each scenario, the missing \acp{RIR} were reconstructed using DiffusionRIR and then integrated into the \ac{MVDR} beamforming framework.

\vspace{2pt}
\noindent \textbf{Performance Metrics:}
Performance was evaluated using the following measures:
\begin{itemize}
    \item \textbf{\ac{SIR}} [\si{dB}], computed separately on speech-only and noise-only segments;
    \item \textbf{\ac{SI-SDR}} [\si{dB}], measured between the clean target signal and the beamformer output;
    \item \textbf{\ac{STOI}}, a perceptual intelligibility score ranging from $0$ to $1$.
\end{itemize}

\vspace{2pt}
\noindent \textbf{Results and Analysis:}
Results for directional and diffuse noise conditions are summarized in Table~\ref{tab:mvdr_virtual_mics_direct_diffuse}. 
We first report baseline performance obtained from the raw microphone signals without beamforming (`Mics'), followed by \ac{MVDR} results using all $16$ ground-truth \acp{RIR} (`Full'). 
For each masking configuration, beamformer performance is compared between two scenarios: using only the measured microphones and their corresponding \acp{RIR} (`Missing'), and using all microphones augmented with the reconstructed \acp{RIR} (`Inpainted').

\begin{table}[t]
\centering
\caption{MVDR results averaged over all SNRs under directional and diffuse noise.}
\label{tab:mvdr_virtual_mics_direct_diffuse}
\resizebox{\linewidth}{!}{%
\begin{tabular}{lccc|ccc}
\toprule
& \multicolumn{3}{c|}{\textbf{Directional noise}} & \multicolumn{3}{c}{\textbf{Diffuse noise}} \\
\textbf{Method}
& \textbf{SIR [dB]}$\uparrow$ & \textbf{SI-SDR [dB]}$\uparrow$ & \textbf{STOI}$\uparrow$
& \textbf{SIR [dB]}$\uparrow$  & \textbf{SI-SDR [dB]}$\uparrow $& \textbf{STOI}$\uparrow $ \\
\midrule
Mics & 0 & -4 & 0.59 & 0 & -4 & 0.58 \\
Full & 12.2 & {9} &{0.88} & 6.2 & {10.2} & {0.79} \\
\midrule
Mask 0 - Missing & 11 & 6 & 0.80 & 4.8 & 7 & 0.70 \\
Mask 0 - Inpainted & {12.3} & {8.8 }& {0.88} &{6.2} & {10.1} & {0.79} \\
\midrule
Mask 1 - Missing & 9.5 & 4.3 & 0.78 & 3.4 & 6 & 0.68 \\
Mask 1 - Inpainted & {12.3} & {8.5} &{0.88}&{6.3}& {10} & {0.78} \\
\midrule
Mask 2 - Missing & 7 & 4.2 & 0.75 & 2 & 5.8 & 0.69 \\
Mask 2 - Inpainted & {12.4} &{7.7} & {0.87} & {6.2}& {8} & {0.77} \\
\midrule
Mask 3 - Missing & 7 & 4.2 & 0.75 & 2 & 5.8 & 0.68 \\
Mask 3 - Inpainted & {12.4} & {4.3} & {0.83} & {5.4}& {5.7} & {0.72} \\
\bottomrule
\end{tabular}%
}
\vspace{-6pt}
\end{table}


The main observations are as follows:
\begin{itemize}
    \item The interpolated \acp{RIR} yield performance close to the full-\ac{RIR} case in terms of \ac{SIR} and \ac{STOI}.
    \item For \ac{SI-SDR}, masks~0 and~1 achieve results comparable to the full-array case, while masks~2 and~3 show degradations of approximately $1$~dB and $4$~dB, respectively.
    \item In all cases, the use of the reconstructed \acp{RIR} substantially outperforms the use of measured microphones alone, indicating that the reconstructed \acp{RIR} closely approximate the real acoustic responses.
\end{itemize}



\subsection{Null Steering and Interference Cancellation}
Now we turn to another task, for which it is required to steer a null towards an interference source. For example, as described in \cite{gannot2017consolidated, gannot2002signal}, the \ac{GSC} structure directs a spatial null to obtain a noise-dominant reference signal, which is then used by the adaptive noise cancellation filters (e.g., implemented via the \ac{LMS} adaptation) to suppress interference and recover the target speech. 
We use the reconstructed \acp{RIR} to impose spatial nulls in prescribed directions and demonstrate that incorporating the reconstructed \acp{RIR} alongside the measured ones results in increased null depth.
Note that, unlike \ac{MVDR}, this analysis does not require estimation of the noise covariance matrix $\boldsymbol{\Phi}_{nn}$; however, the resulting null steering is still applied to the microphone signals.

The evaluation procedure is as follows.
The columns of $\mathbf{H}$ are transformed to the frequency domain to obtain $\mathbf{H}(f) \in \mathbb{C}^{F \times N}$, where $F$ denotes the number of frequency bins. 
For each frequency bin $f$, the corresponding row defines the microphone response vector $\mathbf{h}(f) \in \mathbb{C}^{N \times 1}$ for the ground truth and $\hat{\mathbf{h}}(f)$ for the reconstructed data. 
We compute
\begin{equation}
    \mathbf{d}(f) = \left( \mathbf{I} - \frac{\hat{\mathbf{h}}(f)\hat{\mathbf{h}}^\textrm{H}(f)}{\|\hat{\mathbf{h}}(f)\|^2} \right) \mathbf{h}(f),
\end{equation}
where $\mathbf{I}$ is the identity matrix and $(\cdot)^\textrm{H}$ denotes the conjugate transpose. 
The distance measure,
$\text{Dist}(\mathbf{h}, \hat{\mathbf{h}}) = \sum_{f} \frac{\|\mathbf{d}(f)\|}{\|\mathbf{h}(f)\|}$,
quantifies the alignment between $\mathbf{h}(f)$ and $\hat{\mathbf{h}}(f)$, with lower values indicating better agreement.

As a non-generative baseline, we employ \ac{SCI} \cite{de1978practical}. 
Results are reported in Table~\ref{tab:null_space_ratio} for all masking scenarios.
\begin{table}
    \centering
    \caption{Alignment between interpolated and gound-truth \acp{RIR} for proposed and baseline methods (best results in bold).}
    \label{tab:null_space_ratio}
    \vspace{0.2cm}
    \begin{tabular}{lcccc}
        \hline
        \textrm{Dist($\mathbf{h}$, $\mathbf{\hat{h}}$)}$\downarrow$ & \textbf{Mask 0} & \textbf{Mask 1} & \textbf{Mask 2} & \textbf{Mask 3} \\
        \hline
        Inpainted (Ours)     & \textbf{0.02} & \textbf{0.05} & \textbf{0.19} & \textbf{0.52} \\
        Interpolated (SCI)   & 0.1 & 0.17 & 0.4 & 0.63 \\
        \hline
    \end{tabular}
\end{table}

For mask types 0--2, the reconstructed \acp{RIR} achieve very low distance values, closely matching the ground truth. 
Mask type 3 exhibits larger deviations due to its challenging geometry, where all measured microphones are confined to one side of the array and the missing microphones are spatially distant. 
Nevertheless, even in this case, the proposed method substantially outperforms the \ac{SCI} baseline.


\section{Generalization to Real Data}

In \cite{della2025diffusionrir}, DiffusionRIR was evaluated only on simulated \acp{RIR}.
The current study advances the experimental validation by moving from simulated data to real-world acoustic measurements.

\vspace{1pt}
\noindent \textbf{Dataset:} For real-world evaluation, we used the MeshRIR dataset \cite{koyama2021meshrir}, a large-scale database of densely measured \acfp{RIR}. 
This dataset provides \acp{RIR} sampled on three-dimensional grids, enabling high-spatial-resolution studies. 
The dataset was recorded in a reverberant room with dimensions $7.0 \times 6.4 \times \SI{2.7}{\meter}$ and a reverberation time of approximately $T_{60} \approx \SI{300}{\milli\second}$. 
The loudspeaker was positioned approximately $\SI{2}{\meter}$ from the array plane. Recordings were acquired at a sampling rate of $\SI{48}{\kilo\hertz}$ and subsequently downsampled to $\SI{8}{\kilo\hertz}$ to align with previous experiments. Each \ac{RIR} was truncated to $2048$ samples ($\SI{256}{\milli\second}$), capturing the direct path and early reflections.

\begin{figure}[h!]
 \centerline{\framebox{
 \includegraphics[width=7.8cm]{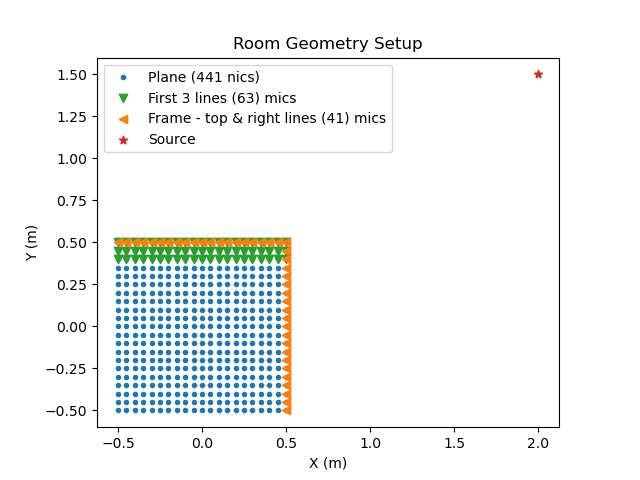}}}
 \caption{Geometric setup of the room with a source and the microphone array, taken from the MeshRIR dataset.}
 \label{fig:mesh_rir_room_setup}
\end{figure}

We focused on the \textit{S1-M3969} subset of MeshRIR, comprising $21 \times 21 \times 9$ microphone grid (a total of $3969$ positions) spanning a cube of size $1 \times 1 \times \SI{0.4}{\meter}$. 
From this grid, we extracted a horizontal plane of $441$ microphones ($21 \times 21$) and designed two array configurations for evaluation:
\begin{enumerate}
    \item \textbf{Three Rows:} 63 microphones corresponding to the first three parallel rows of the grid.
    \item \textbf{Frame:} 41 microphones located along the top and left edges of the plane, forming a L-shaped boundary.
\end{enumerate}
These configurations are illustrated in Fig.~\ref{fig:mesh_rir_room_setup}. 

For each setup, we applied random masking by removing a subset of microphones, thereby simulating missing measurements. 
The remaining microphones were used as inputs to DiffusionRIR to reconstruct the missing responses. 
This design allows us to systematically evaluate reconstruction performance under different spatial sampling patterns. 

\vspace{4pt}
\noindent \textbf{Evaluation Metrics:} To assess the quality of reconstructed \acp{RIR}, we adopt two complementary measures. 
The first is the \ac{NMSE} in $\si{\decibel}$, defined as \cite{zea2019compressed}:
\begin{equation}
\text{NMSE}(\mathbf{H}, \hat{\mathbf{H}}) = 10 \log_{10} \left( \frac{1}{L} \sum_{i=1}^{L} \frac{\|\hat{\mathbf{h}}_i - \mathbf{h}_i\|^2}{\|\mathbf{h}_i\|^2} \right),
\end{equation}
where $\mathbf{h}_i$ is the ground-truth response of the $i$th missing microphone, and $\hat{\mathbf{h}}_i$ is its reconstruction. 
The second measure is the \ac{CD}, which evaluates the angular mismatch between true and estimated responses \cite{morgan1998evaluation}:
\begin{equation}
\text{CD}(\mathbf{H}, \hat{\mathbf{H}}) = \frac{1}{L} \sum_{i=1}^{L} \left( 1 - \left( \frac{\mathbf{h}_i^\top \hat{\mathbf{h}}_i}{\|\mathbf{h}_i\| \|\hat{\mathbf{h}}_i\|} \right)^2 \right).
\end{equation}
A \ac{CD} value close to zero indicates strong alignment, while values near one reflect large discrepancies. 

As a non-generative baseline, we employ the classical \ac{SCI} method \cite{de1978practical}, applied directly to the missing channels. 

\vspace{2pt}
\noindent \textbf{Results:} We now present the evaluation of DiffusionRIR on real-world MeshRIR measurements. 
The performance is assessed under varying masking ratios, comparing reconstructed responses with the corresponding ground-truth \acp{RIR}. 

\begin{figure}[htb]
    \centering
    \begin{subfigure}{0.48\linewidth}
        \includegraphics[width=\linewidth]{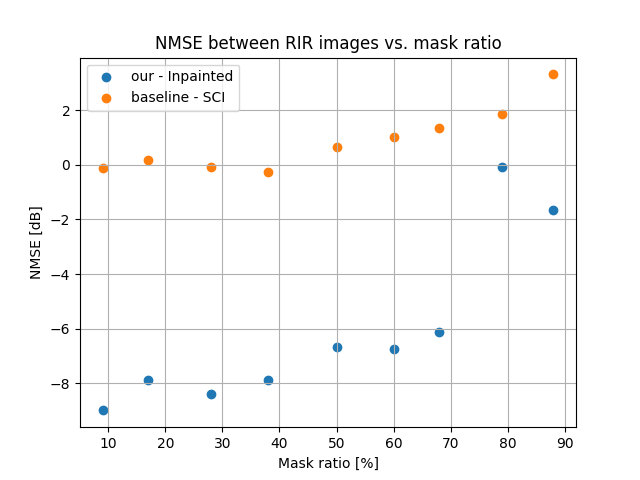}
        \caption{\ac{NMSE}}
    \end{subfigure}
    \hfill
    \begin{subfigure}{0.48\linewidth}
        \includegraphics[width=\linewidth]{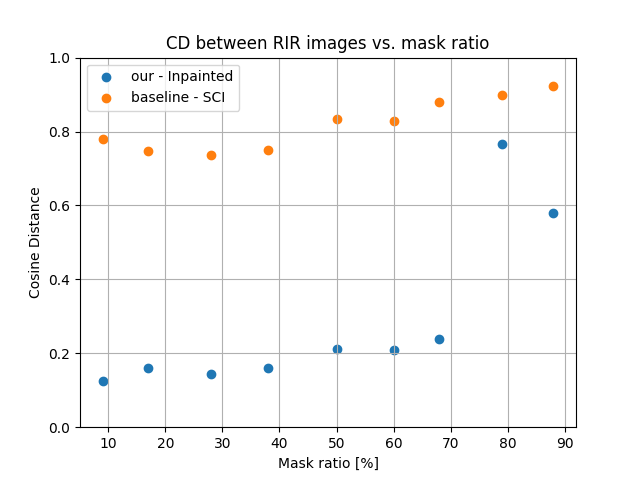}
        \caption{\ac{CD}}
    \end{subfigure}
    \addtolength{\belowcaptionskip}{-2pt} \caption{\ac{NMSE} and \ac{CD} vs. mask ratio for Three Rows configuration.}
    \label{fig:results_3lines}
\end{figure}

\begin{figure}[htb]
    \centering
    \begin{subfigure}{0.48\linewidth}
        \includegraphics[width=\linewidth]{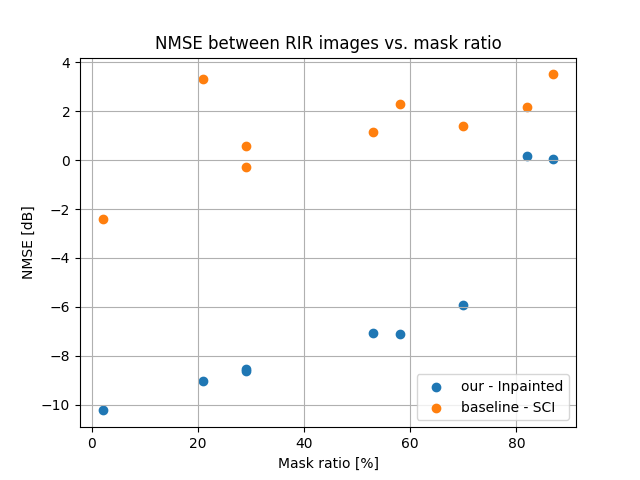}
        \caption{\ac{NMSE}}
    \end{subfigure}
    \hfill
    \begin{subfigure}{0.48\linewidth}
        \includegraphics[width=\linewidth]{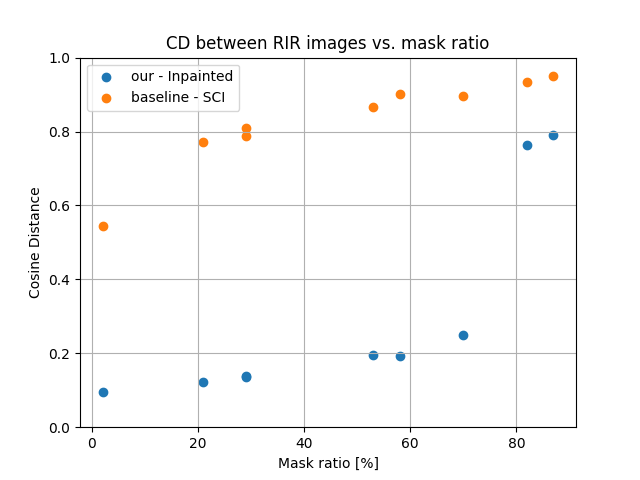}
        \caption{\ac{CD}}
    \end{subfigure}
    \addtolength{\belowcaptionskip}{-2pt} \caption{\ac{NMSE} and \ac{CD} vs. mask ratio for Frame configuration.}
    \label{fig:results_frame}
\end{figure}

Figures~\ref{fig:results_3lines} and \ref{fig:results_frame} summarize the \ac{NMSE} and \ac{CD} as a function of the masking ratio for the \textit{Three Rows} and \textit{Frame} configurations.

%

In both setups, the proposed method significantly outperforms the \ac{SCI} baseline, achieving substantial gains even at high masking ratios.
Although performance degrades as the number of available microphones decreases, DiffusionRIR consistently surpasses standard interpolation methods.

\begin{figure}[ht]
 \centerline{\framebox{
 \includegraphics[width=7.8cm]{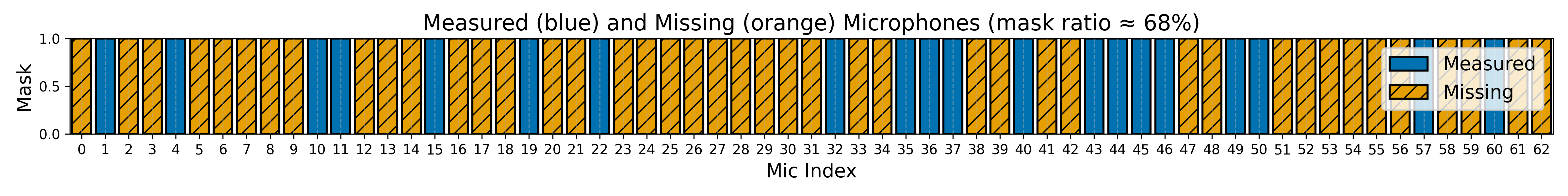}}}
 \addtolength{\belowcaptionskip}{2pt} \caption{Three Rows array configuration: Measured (blue) and missing (orange) microphones.}
 \label{fig:missing_location_3_rows}
\end{figure}

To provide a concrete illustration, we focus on the \textit{Three Rows} setup with a $70\%$ mask ratio. 
Figure~\ref{fig:missing_location_3_rows} depicts the spatial distribution of measured and missing microphones in this configuration. 

\begin{figure}[htb]
    \centering
    \begin{subfigure}{0.47\linewidth}
        \includegraphics[width=\linewidth]{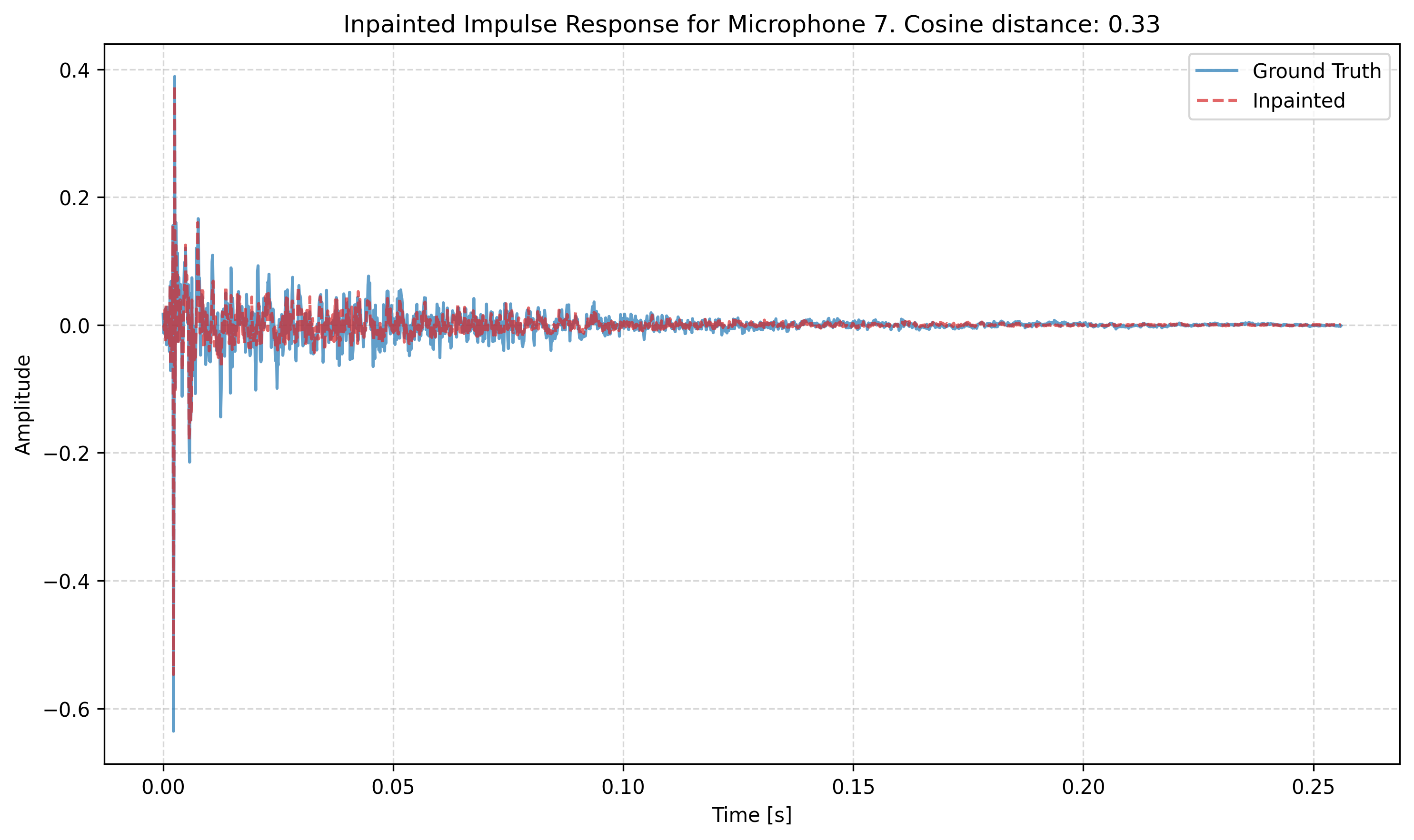}
        \caption{Reconstructed and ground-truth \acp{RIR}}
    \end{subfigure}
    \hfill
    \begin{subfigure}{0.48\linewidth}
        \includegraphics[width=\linewidth]{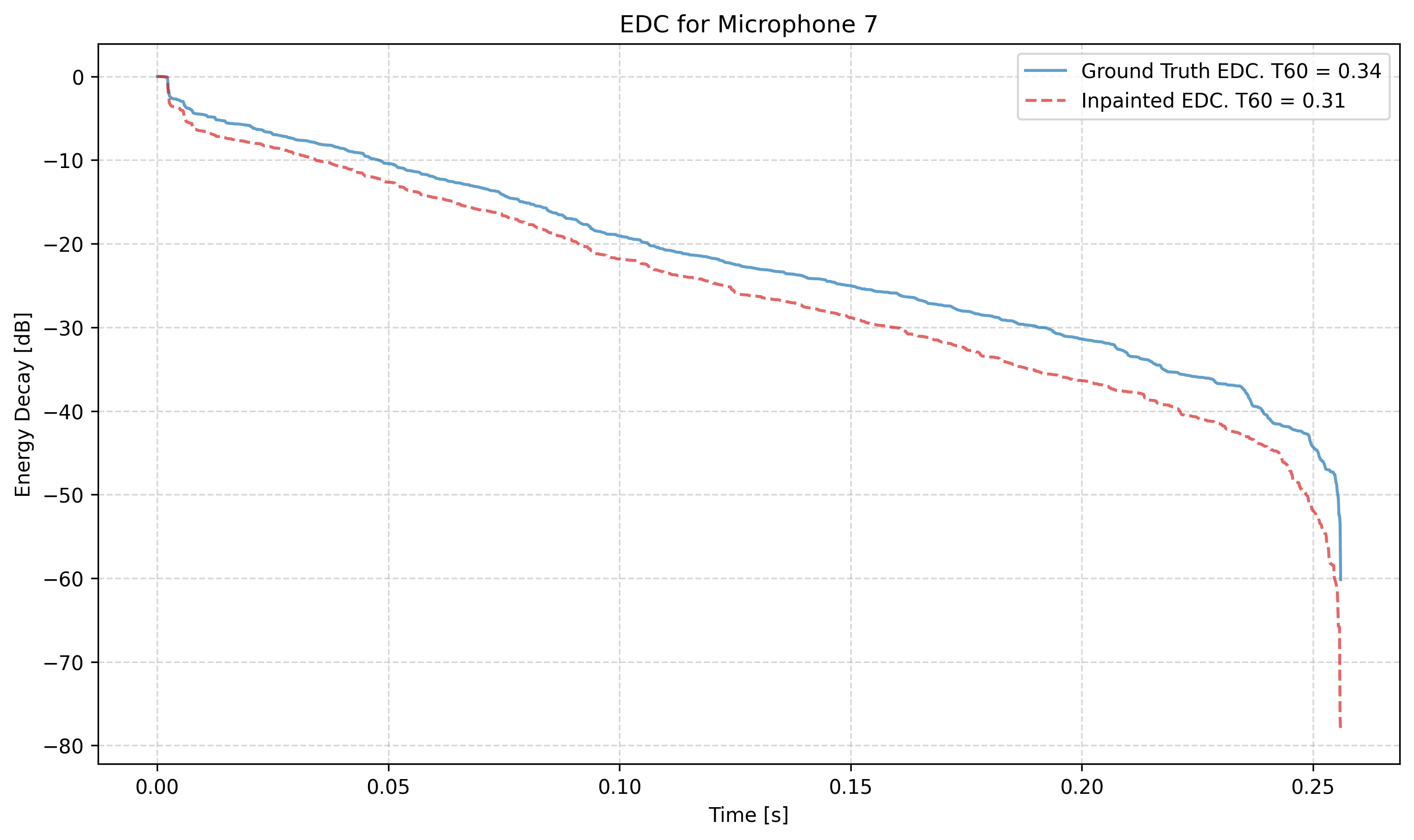}
        \caption{\ac{EDC} and estimated $T_{60}$ values}
    \end{subfigure}
     \addtolength{\belowcaptionskip}{0pt} \caption{Comparison between reconstructed and ground-truth impulse responses and \acp{EDC} for microphone \#7.}
    \label{fig:rir_edc_7}
\end{figure}

We examine the reconstructed response, exemplified for microphone \#7, positioned in a region with relatively sparse neighboring measurements. 
Figure~\ref{fig:rir_edc_7} depicts both the reconstructed and ground-truth \acp{RIR} as well as their corresponding \acp{EDC}. 
The reconstruction preserves the main acoustic characteristics, including the direct path and early reflections. 
The \ac{EDC} curves exhibit similar decay slopes, yielding $T_{60}=\SI{0.31}{\second}$ for the reconstruction and $T_{60}=\SI{0.34}{\second}$ for the ground truth, indicating good preservation of the room’s reverberation characteristics.



These results confirm that DiffusionRIR generalizes from simulation to real data, providing reliable reconstructions under severe sparsity and outperforming traditional interpolation in both error and perceptual metrics.

\section{Discussion}



We addressed the problem of reconstructing \acp{RIR} at unmeasured microphone positions and evaluated DiffusionRIR on real-world acoustic measurements.
The results demonstrate strong transfer from simulated to measured environments, while preserving spatial structure and reverberation characteristics.
These findings indicate that DiffusionRIR captures underlying acoustic structure beyond idealized simulation assumptions and remains effective on real-life data.

Building on this validation, we integrate the interpolated \acp{RIR} into classical microphone array processing tasks and show that their incorporation yields significant performance improvements.


\bibliography{bibliography}

\end{document}